\documentstyle[12pt,aaspp4]{article}
\def\lsim{\lower.5ex\hbox{$\; \buildrel < \over \sim \;$}}
\def\gsim{\lower.5ex\hbox{$\; \buildrel > \over \sim \;$}}
\input psfig.tex
\lefthead{Mukhopadhyay}
\righthead{Stability of Disks Under Nucleosynthesis}

\def\lsim{\lower.5ex\hbox{$\; \buildrel < \over \sim \;$}}
\def\gsim{\lower.5ex\hbox{$\; \buildrel > \over \sim \;$}}

\begin{document}

\title{Stability of Accretion Disks in Presence of Nucleosynthesis}
 
\author{Banibrata \ Mukhopadhyay$^{1,2}$ and Sandip K.\ Chakrabarti$^{1,3}$}

\affil{\small 1. S.N. Bose National Centre for Basic Sciences,\\
JD-Block, Sector III, Salt Lake, Calcutta 700098, India\\
2. Theory Group, Physical Research Laboratory, Navarangpura,\\
Ahmedabad-380009\\
3. Centre for Space Physics, 114/v/1A Raja S.C. Mallick Rd., Calcutta-700047\\
bm@boson.bose.res.in and chakraba@boson.bose.res.in}

\begin{abstract}
We study the effect of nuclear reaction on a thin, axisymmetric, 
differentially rotating, inviscid, steady accretion flow around a black hole
from an analytical point of view. We find that for most of 
the reasonable disk parameters, when $p-p$-reaction, 
dissociation of deuterium and helium are taken into account, 
the {\it transonic} region of the disk continues to have the inner 
sonic point and if the temperature of the flow at the injection point could be
raised (by say, some heating processes) the flow would to pass 
through this inner sonic point. Otherwise, the flow may be unstable.
We use the sonic point analysis to study the solution.
In the rest of the disk parameters the inner sonic point 
is absent altogether and the flow will definitely be unstable. 
\end{abstract}


\section{Introduction}

Black hole accretion flow can reach a temperature of about $T_{virial}=
\frac{1}{k}\frac{GM_{BH} m_p}{r_g} \sim 5.2 \times 10^{12}$K if no cooling processes
are included. Here, $r_g=2GM_{BH}/c^2$ is the gravitational radius of a black hole,
$k$ is the Boltzmann constant, $G$ is the Gravitational constant, $M_{BH}$ is the mass
of the black hole, $m_p$ is the mass of the proton and $c$ is the velocity of 
light.  In a realistic flow, the temperature is smaller. For instance, in a 
Shakura-Sunyaev (1973) Keplerian disk, the temperature becomes close to a few 
times $10^7$K. In a two-temperature transonic flow,
the ions may remain hot ($T_p \sim 10^{8-11}$K) while the electrons may be cooler 
(($T_e \sim 10^{7-9}$K) depending on accretion rate of the Keplerian 
and sub-Keplerian components (Colpi, Maraschi \& Treves, 1984;
Chakrabarti \& Titarchuk, 1995). Since temperature inside a star is
much less compared to the ion temperature mentioned above, considerable
nucleosynthesis is expected inside an accretion disk. It is true that
the free-fall timescale $t_{ff}\sim {r}{v_{ff}} \sim (\frac{r}{r_g})^{3/2} 
\frac{2GM_{BH}}{c^3}$ s is very small compared to the stellar age and density 
of gas inside an accretion disk, $\rho_{d} \sim \frac{{\dot M}_{Ed} t_g} 
{r_g^3} (\frac{r}{r_g})^{-3/2}$gm cm$^{-3}$ is also very small compared 
to the stellar density but the reaction cross-sections are often very 
temperature sensitive ($\propto T^{4-16}$) and the higher temperature 
could compensate for the lower density.

Considerable work has been done in the past on nucleosynthesis  inside a disk
taken into consideration. Paczy\'nski and Jaroszy\'nski (1978) have 
studied the importance of nucleosynthesis at high temperature thin disks
made up of Helium and Carbon. In these disks, electron degeneracy pressure
supports gravity. Taam \& Fryxell (1985) studied thin accretion disks 
of hydrogen-rich matter and showed that under certain
condition, thermal instability of a Keplerian disk may totally disappear. They 
consider very high density disks where degenerate pressure in the mid-plane
supports the disk structure. Chakrabarti, Jin \& Arnett (1987) and Jin, Arnett
\& Chakrabarti (1989) studied nucleosynthesis in thick accretion disks
of realistic density and temperature. These disks have very little radial velocity
and a significant amount of nucleosynthesis was possible only if
viscosity parameter of Shakura-Sunyaev (1973) is very low $\alpha_{ss} \lsim
10^{-4}$. They showed that in order that {\it thin} disks remain stable against
convection, only $p-p$ reactions are allowed in these disks in the radiation
pressure supported regions. However, for hot thick disks, even CNO cycle would be
allowed without destabilizing the disk. These results were later repeated by 
independent groups (Arai \& Hashimoto, 1992; Hashimoto et al. 1993). 
More recently, Chakrabarti \& Mukhopadhyay (1999) and Mukhopadhyay \& Chakrabarti 
(2000) revisited the problem with a self-consistent accretion disk solution which
includes heating and cooling. They find that nucleosynthesis in some range of 
the parameter space could be high enough to have an energy release or absorption 
by nuclear reaction comparable to the viscous heating. These studies are finding
attention in recent years since one could detect lines from various isotopes
from jets and companions and can, in principle, have some estimate of the
conditions around a black hole.

However, work mentioned above was purely numerical in nature and was mainly due to 
repeated use of two types of analysis: First, the steady state thermodynamic 
condition of a disk is determined by using fourth order Runge-Kutta 
method {\it without} nucleosynthesis taken into account. Using temperature 
and density distributions thus obtained at each point, nucleosynthesis code is run which
mainly involves solving 255 linear equations simultaneously, one equation for 
each isotope. For each infinitesimal advancement of matter toward the black 
hole, nuclear composition was updated and energy release or absorption 
was computed locally. After one complete run, the energy release is 
incorporated into the hydrodynamic equations and the integration scheme using 
the Runge-Kutta method is carried out to have updated 
thermodynamic conditions. This process is continued till
convergence. In the present paper, our approach 
is completely different. We attack the problem analytically and estimate effects
of nucleosynthesis by using sonic point analysis. This way the problem becomes
fully self-consistent to begin with and no iterative step is needed. Presence of
a saddle type sonic point is a necessary condition for a steady solution, but it
is not sufficient. The flow must connect a black hole horizon with infinity.
Of course, in order to have the problem under control, we chose only a few reactions
which are likely to be very important, namely, the Proton-Proton 
and photo-dissociation of deuterium and helium. Since we are interested in stability,
we considered only those reactions (both exothermic and endothermic)
which change specific energy (either way) significantly. We, however,
neglect viscous heating or possible radiative transfer effects. 

In any successful model of the black hole accretion, a source of the soft photon 
and a hot, Comptonising cloud must be present. In our model, this Comptonising cloud 
is the inner part of the disk itself. For concreteness, we concentrate on this model 
of black hole accretion, namely, one in which the flow close to a black hole is transonic (this is a
necessary condition for any model), but far away, it is sub-divided into
Keplerian and sub-Keplerian components. According to the properties of the transonic disks
(see, Chakrabarti \& Titarchuk, 1995; and references therein) higher viscosity flow settles
into the equatorial plane, while quasi-spherical flows with lower viscosity 
surround the Keplerian disk. Soft photons emitted from the Keplerian component are
reprocessed by the electrons at the centrifugal pressure supported boundary layer and hard X-rays
are emitted. Recent observational results strongly support the presence of 
Comptonising clouds close to a black hole (Mirabel \& Rodriguez, 1999;
Homan et al., 2000) which have properties very similar to our 
centrifugal pressure supported regions in the inner accretion disks. 

We had to choose a particular model in order to compute the optical depth of matter
through which soft photons are Comptonized. Since at high temperature, photo-dissociation
is important one has to compute number of hard photons present in this inner disk.
Ordinarily, photo-dissociation at a high temperature is taken for granted, but in the
present situation, there are paucity of very hot photons since only few of them
are emitted from the Keplerian component and even fewer are intercepted by the
sub-Keplerian flow located ahead of the Keplerian disk. What is more, optical depth may be
so low that most of the time these photons do not interact adequately with 
deuterium and helium.  Because of this, we compute photon number density 
self-consistently from the emitted spectrum of a sub-Keplerian flow. 

In the next Section, we present the basic hydrodynamic equations which 
govern the steady state flow and perform the sonic point analysis.
We show that effect of the nuclear term is significant in changing 
topological properties of the flow. In \S 3, we show 
that number of sonic points is larger in presence of nucleosynthesis.  
We present a complete global solution and show how it is modified in 
presence of nucleosynthesis. Finally, in \S 4, we draw our conclusions.

\section{Model Equations}

In what follows, we use the unit of distance (x) to be $r_g=2GM_{BH}/c^2$, 
unit of velocity to be $c$ and the unit of mass to be $M_{BH}$, where $G$,
$M_{BH}$ and $c$ are the gravitational constant, mass of the black hole and
the velocity of light respectively. We use Paczy\'nski \& Wiita (1980)
potential $\phi(x)=-\frac{1}{2(x-1)}$ to describe the flow around a 
Schwarzschild black hole. The basic dimensionless hydrodynamic equations 
which govern the infalling matter in the steady state are given by:

\noindent (a) The radial momentum equation:

$$
\vartheta \frac{d\vartheta}{dx} +\frac{1}{\rho}\frac{dP}{dx}
+\frac {\lambda_{Kep}^2-\lambda^2}{x^3}=0,
\eqno{(1a)}
$$

\noindent (b) The continuity equation:

$$
\frac{d}{dx} (\Sigma x \vartheta) =0 ,
\eqno{(1b)}
$$

\noindent (c) The entropy equation:

$$
\frac{2na{\rho}{\vartheta}h(x)}{\gamma}
\frac{da}{dx}-\frac{a^2{\vartheta}h(x)}
{\gamma}\frac{d{\rho}}{dx}=Q_{nuc} .
\eqno{(1c)}
$$
We do not have an azimuthal momentum equation, since we assumed 
an inviscid flow where angular momentum is conserved. The sound speed 
is defined by, $a^2=\frac{\gamma P}{\rho}$, where, $P$ and $\rho$ are the 
pressure and the density respectively and $\gamma$ is the adiabatic index. 
Also, $\lambda$ is the conserved specific angular momentum (in units of $2GM_{BH}/c$) 
of the infalling matter, $\lambda_{Kep}$ is the specific angular momentum in a 
Keplerian disk, $\lambda_{Kep}^2 = \frac{x^3}{2(x-1)^2}$, $\Sigma$ is the 
vertically integrated density, $h(x)$ is the half thickness of the disk 
[$\sim a x^{1/2}(x-1)$], $n = \frac{1}{\gamma-1}$ is the polytropic index, 
$Q_{nuc}$ is the height averaged heat generation/absorption due to 
the nuclear effect of the disk. In order to isolate the effect 
of nucleosynthesis alone, we neglect all the other heating/cooling terms. The 
factor $Q_{nuc}$ may be positive or negative depending 
on whether net nuclear energy is locally  exothermic or endothermic. 
In future, we plan to include viscous heating and radiative cooling terms.

The energy release rate for $p-p$ reaction is given by,
$$
Q_{pp}=\frac{\rho_{p}^2}{2}<\sigma v>_{pp} q_{pp} ,
\eqno{(2)}
$$
where, $\rho_{p}=\rho N_{Av} X_p/A_p$ is the number density of protons.
The energy absorption rate due to $D$ and $^4\!He$ dissociation is,
$$
Q_{disso}=\rho_i <\sigma v>_i q_i,
\eqno{(3)}
$$
where, the number density of the $i$th species (proton, deuterium or helium for this case) is  
$\rho_i={\rho N_{Av} X_i}/{A_i}$. $N_{Av}$ is the Avogadro Number, $X_i$ is the mass abundance
and $A_i$ is the mass number of the $i$th element 
and $q_i$ is the Q-value of the corresponding reaction. The reaction 
rates have been taken from standard literature (e. g., Clayton, 1983). For simplicity, and 
in order that this work may be generalized even when a large number of nuclear species are 
considered, we can express this as,
$$
<\sigma v>_i=exp(g_i),
\eqno{(4)}
$$
where, 
$$
g_i=c^1_i+c^2_i/t9+c^3_i/t9^{1/3}+c^4_i t9^{1/3}+c^5_i t9 +c^6_i t9^{5/3}+c^7_i log(t9)
$$ 
is a seven parameter family of functions (Thielemann, 1980). For different reactions,
the constant coefficients $c_i$s will be different. In our case, $Q_{nuc}$ is the 
dimensionless height integrated nuclear energy generation rate $Q_{nuc}=(Q_{pp}+Q_{d}+Q_{He}) h(x)$.

Using Eqs. (1b-1c), we eliminate all $\frac{d\rho}{dx}$ and $\frac{da}{dx}$ terms and get, 
$$
\frac{d\vartheta}{dx}=\frac{\frac{(\gamma-1)}{\vartheta}(Q_{pp}+Q_d+Q_{He})+
(\gamma+1)[\frac{1}{2(x-1)^2}-\frac{\lambda^2}{x^3}]-
\frac{a^2(5x-3)}{x(x-1)}}{\frac{2a^2}{\vartheta}-(\gamma+1)\vartheta}.
\eqno{(5)}
$$
At the sonic point, numerator and denominator both simultaneously should be zero, so,
$$
\frac{2a^2}{\vartheta}-(\gamma+1)\vartheta=0
\eqno{(6)}
$$
and
$$
\frac{(\gamma-1)}{\vartheta}(Q_{pp}+Q_d+Q_{He})+
(\gamma+1)[\frac{1}{2(x-1)^2}-\frac{\lambda^2}{x^3}]-
\frac{a^2(5x-3)}{x(x-1)}=f(a,\vartheta)=0.
\eqno{(7)}
$$
From (6), we can get the Mach number at the sonic point as,
$$
M_c=\sqrt{\frac{2}{\gamma+1}}.
\eqno{(8)}
$$
From (7) and (8) we can obtain the sound speed at the sonic point.

The energy and entropy at the sonic points are expressed as,
$$
E_c=\frac{1}{2}v_c^2+na_c^2-\frac{1}{2(x_c-1)}+\frac{\lambda_c^2}{2x_c^2} ,
\eqno{(9)}
$$
$$
{\dot{\cal M}}=a_c^{(2n+1)}{x_c}^{3/2}{(x_c-1)}{v_c} .
\eqno{(10)}
$$

Unlike the case of a flow with conserved energy (Chakrabarti, 1989; hereafter C89)
one cannot compute everything about the flow using only two parameters,
i.e., energy and specific angular momentum. In the present context of 
nuclear energy generation, one requires to supply  the accretion rate as
well which in turn determines the density of the flow. This is explicitly needed
in eq. (7) above.

During the infall of matter, rate of change of mass fraction for proton, 
deuterium and helium can be written as,
$$
-\frac{dX_{p}}{dt}=-\vartheta\frac{dX_{p}}{dx}
=X_{p}exp(g_{pp})\rho-\frac{X_dexp(g_{d})}{2},
\eqno{(11)}
$$
$$
-\frac{dX_{d}}{dt}=-\vartheta\frac{dX_{d}}{dx}
=X_dexp(g_d)-\frac{X_{He}exp(g_{He})}{2}-X_{p}
exp(g_{pp})\rho,
\eqno{(12)}
$$
$$
-\frac{dX_{He}}{dt}=-\vartheta\frac{dX_{He}}{dx}
=X_{He}exp(g_{He}).
\eqno{(13)}
$$
During the Runge-Kutta integration, these linear differential equations 
are also solved simultaneously to obtain their mass fractions
at each point.

\section{Calculation of Photon Number Density}

Before we present results of our global analysis, we 
like to spend some time on how the photon number density 
is computed inside the transonic flow. Usual treatment of 
photo-dissociation in the literature assumes the presence of a
black body radiation in the flow. However, advective flow 
close to a black hole is likely to be optically thin,
and they would intercept soft photons from Keplerian disks located
farther out. These soft photons are Comptonized and in reality, 
the photo-dissociation would be done by these Comptonized photons
and the synchrotron photons rather than black body photons. We need 
to compute the number of hard photons above $2$ MeV in order 
to check if this is much larger compared to the deuterium nuclei.

If one considers a pure black body spectrum then the photon 
number density can be written as,
$$
N_\nu^{BB}=\frac{8\pi}{c^3}\left(\frac{kT}{h}\right)^3\int^{\infty}_{x_m}\frac{x^2dx}{e^x-1}
\eqno{(14)}
$$
where, $x=\frac{h\nu}{kT}$. But the spectrum of photons is not purely blackbody in nature.
To calculate the number density of photon in our system we use results 
of Chakrabarti (1997, hereafter C97). We concentrate on Fig. 4 of C97
where variation of photon spectrum other than black body is discussed. 
In Fig. 4a of that paper, spectral variation is shown with energy for different 
values of Keplerian accretion rate (in units of Eddington rate)
$\dot{m}_d$. For a complete dissociation of deuterium, at least $2$ MeV 
energy is needed (although with the energy of slightly less than $2$ MeV 
dissociation may start.). We first concentrate on those cases of 
$\dot{m}_d$ where curves are extended at least to the region
greater than $2$ MeV. Each of these curves has one peak due to multi-colour 
black body and the other is a hump due to Comptonization. We concentrate
on the hump at higher frequency to compute the number of hard photons.
For the dissociation of a helium nuclei one requires a higher energy ($\sim 32$ MeV).
In Fig. 4a of C97, except for $\dot{m}_d=1.5$ where there is a large peak 
due to the blackbody emission, other cases have significant amount of hard photon.
Along with these photons, we include hard photons due to synchrotron radiation 
as obtained from a standard distribution $\sim I_\nu \sim \nu^{-1.25}$
normalized in a way such that its number is the same as that of the Comptonized 
photons at the `knee' where blackbody distribution  meets the Comptonized power-law
component. This  procedure provides a realistic estimate of  hard photon. In any case,
our main aim is to see if the resulting photon numbers are very large compared to the
number of deuterium or helium nuclei. So the exact number is not important.

Following the treatment of Wagoner (1969), we completely turn off the 
dissociation process if the number of hard photons at a given radius 
is fewer than about a thousand times that of the deuterium. The same
consideration applies for the dissociation of 
helium nuclei as well.

\section{Results}

Here we discuss a few results assuming matter falls towards  a black hole of 
mass $10M_\odot$. We choose the polytropic index $n$ to be $3$. 
We assume that the accretion flow consists of both Keplerian and sub-Keplerian 
matter (Chakrabarti \& Titarchuk, 1995). We assume that these two rates are $\dot{m}_d$ 
and $\dot{m}_h$ respectively and ${\dot m}_d + {\dot m}_h = 2$
for concreteness. We first compute the electron and proton temperatures
in the flow by using two temperature hydrodynamic equations  numerically
(Chakrabarti \& Titarchuk, 1995) for different ${\dot m}_d$ and
${\dot m}_h$ rates and use the resulting (cooler) temperatures to compute 
nucleosynthesis in our analytical code. Carrying out of nucleosynthesis 
work in presence of Comptonization would be otherwise 
very much prohibitive. We verify a posteriori from the
solution with nucleosynthesis that such an approximation is valid.
In all the three cases listed below we choose $\lambda=1.6$ and outer edge
of the flow at $x_o=1149$. A flow without nucleosynthesis has a smooth solution
passing through the sonic point at $3.05$ with these parameters provided it is
launched at the outer edge at $x_o$ with a radial velocity $v_o=8.8566 \times 10^{-5}$
and sound speed $a_o = 0.061966$.

\subsection{Case I}

Here we choose $\dot{m}_d=0.05$, $\dot{m}_h=1.95$. In this case,  
the photon number density is $1.61\times10^{17}$/cc.
We first check the global behaviour of the infalling matter, such as the 
position and nature of the sonic points if nucleosynthesis is turned on.  Following C89, 
we plot in Fig. 1a, energy ${\cal E}$ at the sonic point as a function of the entropy 
accretion rate ${\dot {\cal{M}}}$ at these points for a fixed 
specific angular momentum ($\lambda=1.6$) of the sub-Keplerian and Keplerian mixture. 
Dashed curve indicates the behaviour when nucleosynthesis is not included. This 
behavior is similar to that observed in C89. The solid curve indicates the variation of energy 
with entropy accretion rate at the sonic point when nucleosynthesis in the 
flow is included. The dashed curve is shifted vertically by $0.01$ in order to 
see the effects of nucleosynthesis prominently. If the ratio of the  number density of photon and 
the nuclear species to be dissociated in the disk is less than $10^3$ 
the dissociation is turned off because of the deficiency of photons (Wagoner, 1969). 
For the present set of parameters, it is seen that till
$\dot{\cal{M}}\sim10^{-5}$ nucleosynthesis does not affect
the sonic points. In the range $\dot{\cal{M}}\sim10^{-5}-10^{-6}$,
the outer sonic points disappear as the solid curve shows a distinct gap in that range 
where nucleosynthesis is prominent. Subsequently, a new branch is originated.
As the inner sonic point is not affected, we can conclude that in these parameters 
the stability of the disk is not affected. Matter can
pass through either the inner sonic point (I) and or
through the outer sonic point  (O) if the location  of the latter 
point is far enough. Flows with sonic points located 
at the intermediate distances are unstable.

In Fig. 1b, we show the variation of the Mach number as a function of
radial distance. The inner sonic point is chosen at $x_{\rm in}=3.05$.
The velocity of the matter at the outer edge of the disk 
(assumed at $1149$) is chosen to be $v_o=8.8566\times10^{-5}$ as this 
allows a smooth solution through the inner sonic point when nuclear 
energy is not included. The long-dashed curve is the resulting solution.
The sound speed at the launching point is $a_o=0.061966$.
All quantities are expressed in geometric units. When nucleosynthesis is 
included, in order that the flow passes through the same inner sonic point 
smoothly, we had to raise the sound speed to be $a_o= 0.0791819$. 
The solid curve indicates this solution.  If the flow with nucleosynthesis 
is given the same $a_o=0.061966$ as that of the smooth solution
without nucleosynthesis (long dashed curve), the resulting solution becomes totally sub-sonic
and is unphysical (dash-dotted curve) in the context of black hole 
accretion. For comparison, we present two other solutions (short dashed 
and dotted) which begin with different sound speed, and are also found to 
deviate from the true transonic solution (solid curve). Fig. 2 shows 
the variation of the energy release due to nucleosynthesis
as a function of the logarithmic radial distance of 
the flow (solid curve). The energy release is basically 
due to $p-p$ reaction as deuterium burning took place instantaneously
after launching the flow. Note that the release is at  the most $10^{17}$ ergs/gm/sec
which the specific energy of the flow with an inner sonic point could be around $0.001c^2$ergs/gm.
Thus the nuclear cooling time scale is $\sim 10$s which is large compared to,
infall or dynamical time scales $t_{d} \sim 10^{-3}$s. Thus, energetically 
nuclear energy release is not significant.
 
\subsection{Case II}

Here, we choose a comparatively hotter flow. We use $\dot{m}_d=0.001$, $\dot{m}_h=1.999$ 
photon number density is calculated as $3.02\times10^{21}$/cc. The global
behaviour and particular solutions are shown in Fig. 3(a-b). 
They have similar properties as those of Case I. In Fig. 3a, we clearly see the 
disappearance of the outer sonic points between $O_1$ and $O_2$. The deviation from 
a `non-nuclear' solution is higher, and one requires to launch the flow with a higher
speed of sound: $a_o=0.082167$. In this case, the gas has to be heated 
at a much higher temperature in order that it may pass through the same sonic point.
In that sense the flow is prone to become more unstable than that of Case I.
Fig. 2 (dashed curve) shows energy release for this case. 
A hotter gas releases somewhat higher energy.  

\subsection {Case III}

In this case, we choose a cooler flow so that deuterium is dissociated 
closer to the black hole. We use $\dot{m}_d=0.4$, $\dot{m}_h=1.6$ and
specific angular momentum of the mixture of Keplerian and sub-Keplerian gas is $\lambda=1.6$ as before.
Since accretion rate is large, there are virtually no Comptonized 
hard photons. However, there are synchrotron photons which dissociate
deuterium. These synchrotron photons are computed from a power-law intensity distribution
with a slope of $-1.25$ joining with blackbody spectrum at roughly the same place 
where Comptonized power-law joins. Fig. 2 shows the total energy release 
(dotted upper curve) as well as energy absorption due to deuterium in 
this case (dotted, marked by $Q_D$). As matter comes closer to the black hole,
some deuterium is produced by the $p-p$ reaction which is also instantaneously 
destroyed. This raises energy absorption closer to a black hole. 
Note also that only absolute value of $Q_D$ is drawn here.

Fig. 4 gives the variation of the abundance of deuterium, proton and neutron with logarithmic radial
distance. In this case, deuterium is totally destroyed at around $100$ Schwarzschild radii
causing a rise in the protons and neutrons.

\section{Concluding remarks}

We studied analytically the effects of nucleosynthesis 
on black hole accretion flows of constant angular
momentum. We took only the basic reactions such as $p-p$ reaction producing deuterium and 
destruction of deuterium and helium due to photo-dissociation since (a) it is easier to
handle only a few reactions analytically and (b) these reactions primarily control the
energetics of the flow. Since we were dealing with an optically thin flow,
the radiation could not be treated as a thermal blackbody 
and the photon number density had to be calculated from the actual thermal
Comptonized spectrum and from synchrotron emission from coronal electrons. Photons with energy 
higher than the threshold value were used to destroy deuterium and helium.

Though, the energy released or absorbed were found to be not very 
significant compared to the rest mass of the flow it has a strong influence
on the stability properties at the transonic region. For instance, we find that 
when nucleosynthesis in included, the sound speed of the injected matter must be
higher by at least $25$ to $30$ percent in order that the flow passes through 
inner sonic point. In other words, the temperature of the flow at the injection 
point must be $50$ to $60$ percent higher. If a physical process (such as 
magnetic heating) could not be found to enable this, the flow 
will not be transonic and would be unstable (See, Figs. 1b and 3b). 

So far, in the literature, study of accretion flows around black holes 
had been done using two-step numerical procedure --- pure hydrodynamics is followed by 
pure nucleosynthesis and results of nucleosynthesis is fed back into hydrodynamics.
This procedure was repeated for the whole disk till a reasonable convergence occurs. 
In the present paper, we combined hydrodynamics and nucleosynthesis for the first time and 
studied analytically how nucleosynthesis influences topology of the flow.
In some region of the parameter space, the outer sonic point is found to be absent.
How would a flow behave in this region? It is conjectured that matter 
would try to pass through `some' outer sonic point, becomes supersonic
and subsequently pass through the inner sonic point. Since the
outer sonic point is not present, the searching procedure to the
true sonic point should give rise to an unstable or oscillating solution.
Only a {\it time dependent} numerical simulation can tell if this conjecture 
is correct. This will be attempted in the future.

{}

\begin {figure}
\vbox{
\vskip 0.0cm
\hskip 0.0cm
\centerline{
\psfig{figure=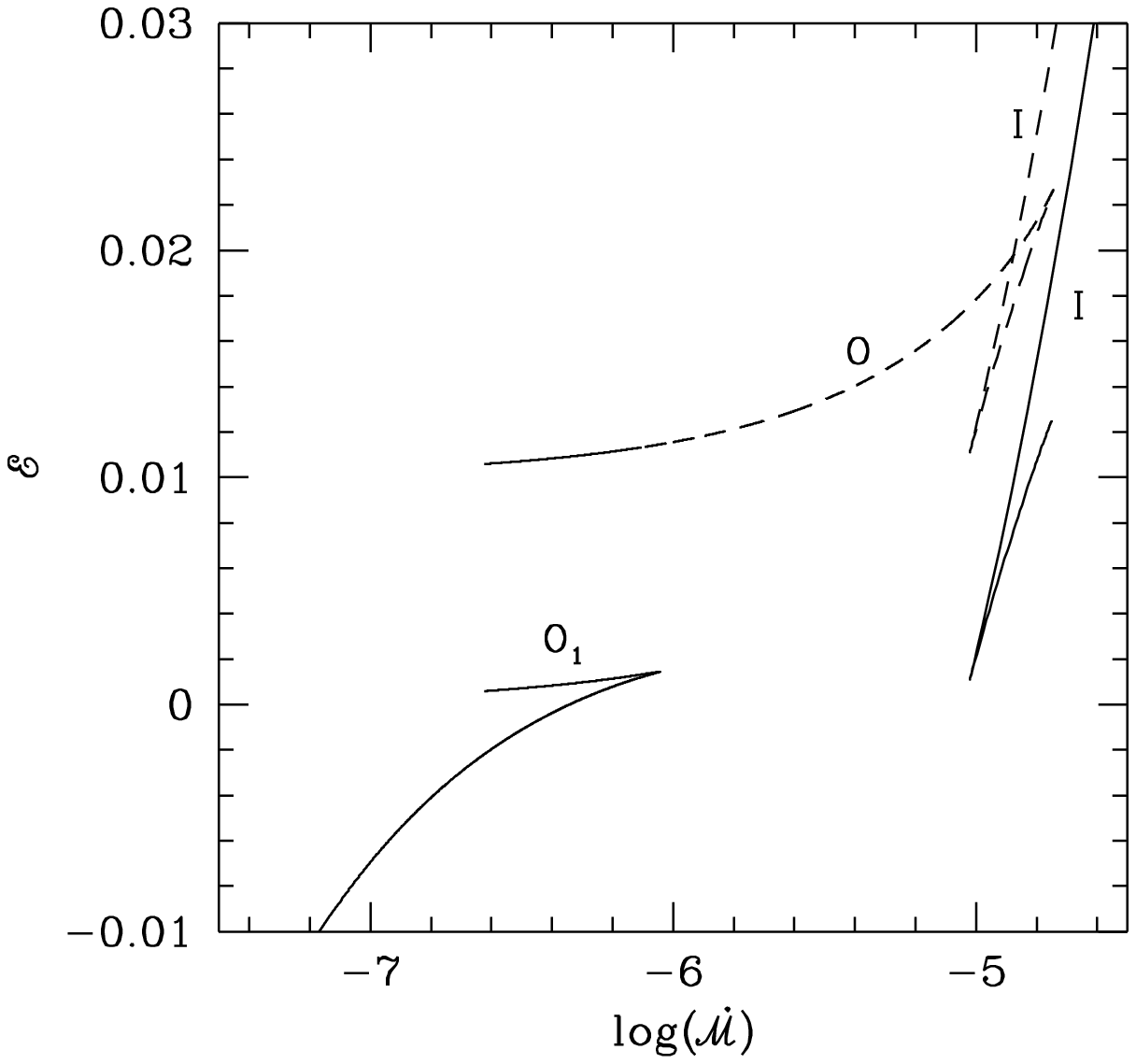,height=8truecm,width=8truecm}}}
\end{figure}
\begin{figure}
\vspace{10.9cm}
\noindent {\small {\bf Fig. 1a}: Variation of the energy with entropy 
accretion rate as the sonic points are changed. I and $O_1$ represent 
the inner and outer saddle type sonic points and the unmarked curves 
represent the unphysical `O' type sonic point. Dashed curve, shifted 
by $0.01$ along energy axis, shows the sonic point behavior
when nucleosynthesis is turned off.}
\end{figure}

\clearpage 

\begin {figure}
\vbox{
\vskip -10.0cm
\hskip 0.0cm
\centerline{
\psfig{figure=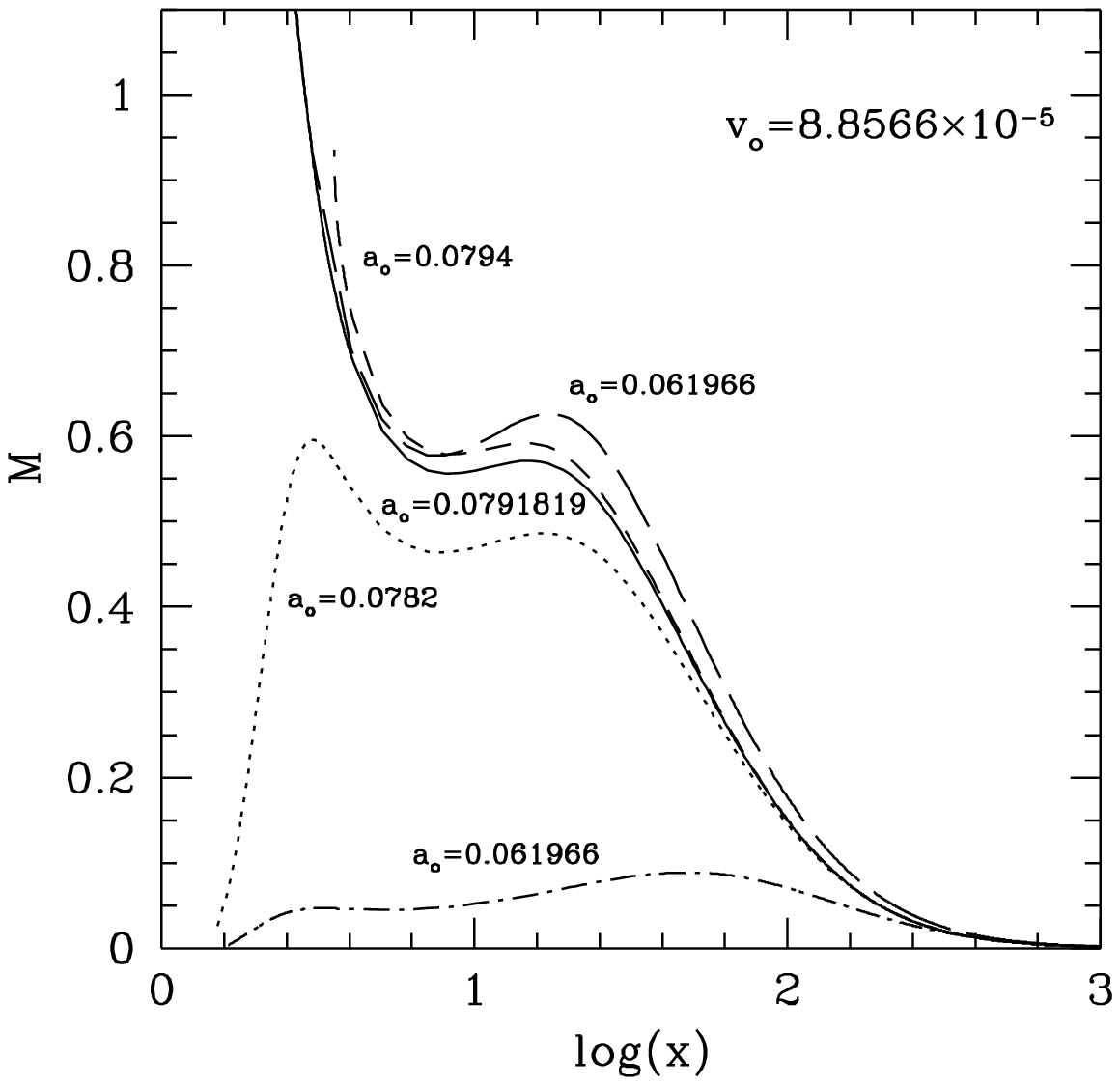,height=8truecm,width=8truecm}}}
\end{figure}

\begin{figure}
\vspace{19.5cm}

\bigskip

\bigskip

\bigskip

\noindent {\small {\bf Fig. 1b} : Variation of the Mach number of the 
flow with logarithmic radial distance. Matter injected at $x=1149$ with 
radial velocity $v_o$ moves in with injected sound speed $a_0=0.061966$ to 
pass through the inner sonic point when nucleosynthesis is turned off (long-dashed 
curve).  When nucleosynthesis is turned on, sound speed must be modified to 
$a_0=0.0791819$ in order that the flow passes through the inner sonic point
(solid curve). Dash-dotted curve is the (unphysical) solution produced 
when nucleosynthesis is included but the flow had the same initial 
parameters as that of the long-dashed curve. short-dashed and 
dotted curves are for some other initial parameters to indicate
non-transonicity. See text for details.}

\end{figure}

\vfill\eject

\begin {figure}
\vbox{
\vskip 4.0cm
\hskip 0.0cm
\centerline{
\psfig{figure=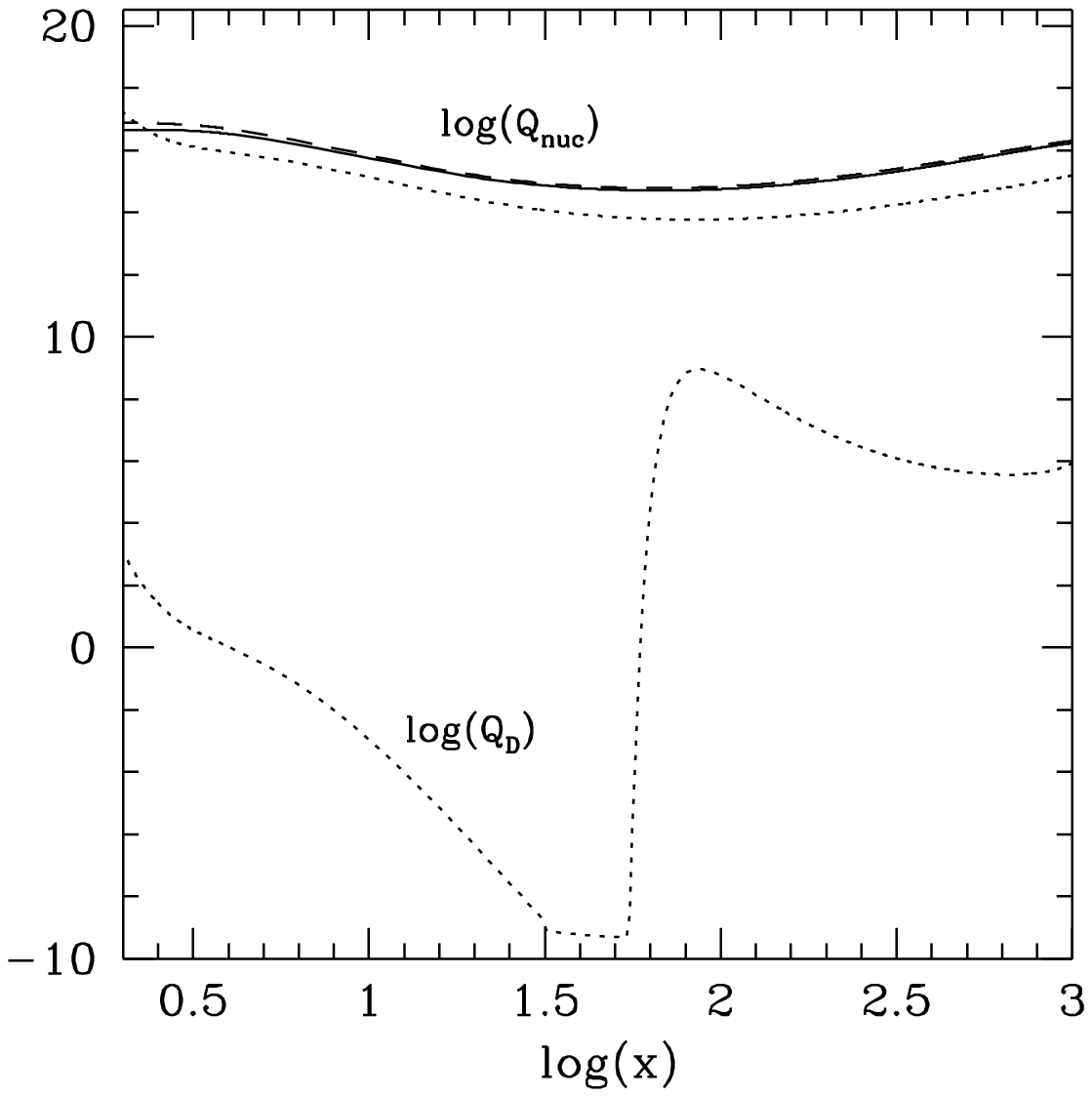,height=10truecm,width=10truecm}}}
\end{figure}
\begin{figure}
\vspace{5.3cm}
\noindent {\small {\bf Fig. 2}:
Variation of net energy release $Q_{nuc}$ (in ergs/gm/sec) as matter 
is accreted on a black hole. Solid and dashed curves are drawn for cases I and
II and the dotted curves are drawn for the case III discussed in the 
text. $Q_D$ is the energy absorption when deuterium is dissociated.}
\end{figure}

\vfill\eject

\begin {figure}
\vbox{
\vskip 4.0cm
\hskip 0.0cm
\centerline{
\psfig{figure=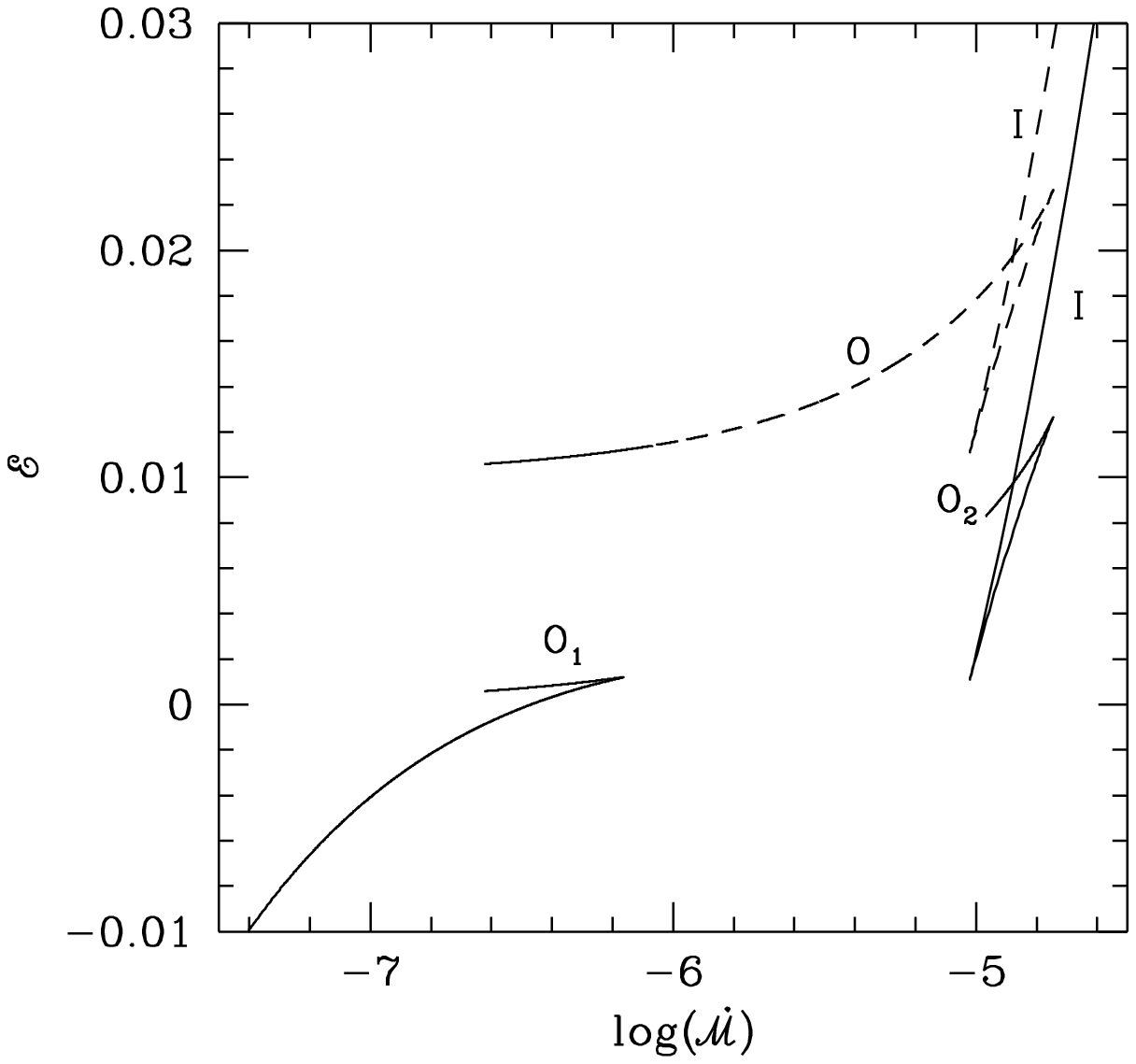,height=8truecm,width=8truecm}}}
\end{figure}
\begin{figure}
\vspace{5.8cm}
\noindent {\small {\bf Fig. 3a}: Same as Fig. 1a except that the flow is 
hotter. $O_1$ and $O_2$ are outer sonic point branches separated by a gap.}
\end{figure}

\vfill\eject

\begin {figure}
\vbox{
\vskip 4.0cm
\hskip 0.0cm
\centerline{
\psfig{figure=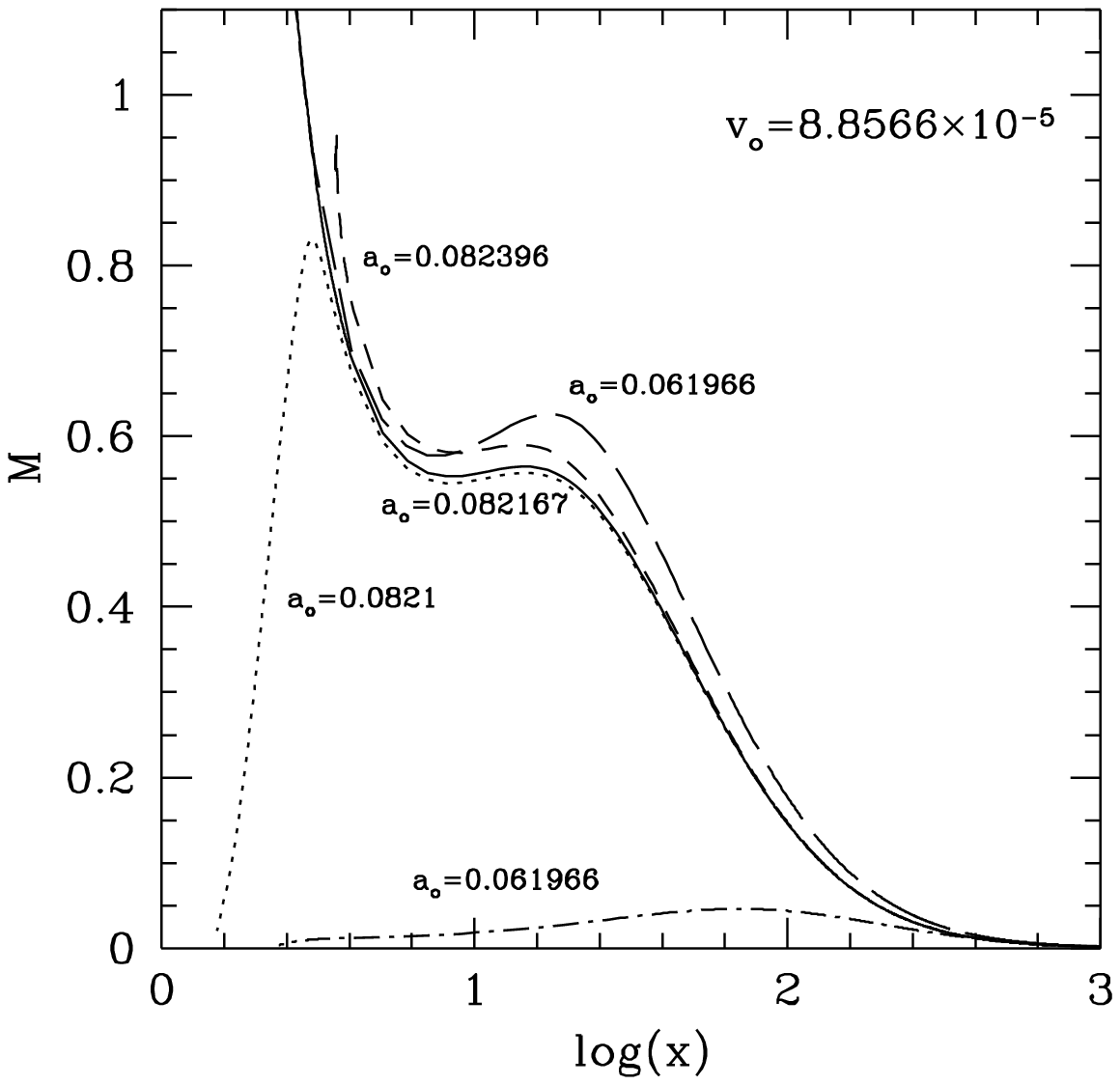,height=10truecm,width=10truecm}}}
\end{figure}
\begin{figure}
\vspace{5.3cm}
\noindent {\small {\bf Fig. 3b}: 
Same as in Fig. 1b, except for a hotter flow.
In this case, when nucleosynthesis is turned on,
the sound speed must be modified to $a_0=0.082167$ in 
order that the flow passes through the inner sonic point (solid curve). 
See text for details.}
\end{figure}

\vfill\eject

\begin {figure}
\vbox{
\vskip 4.0cm
\hskip 0.0cm
\centerline{
\psfig{figure=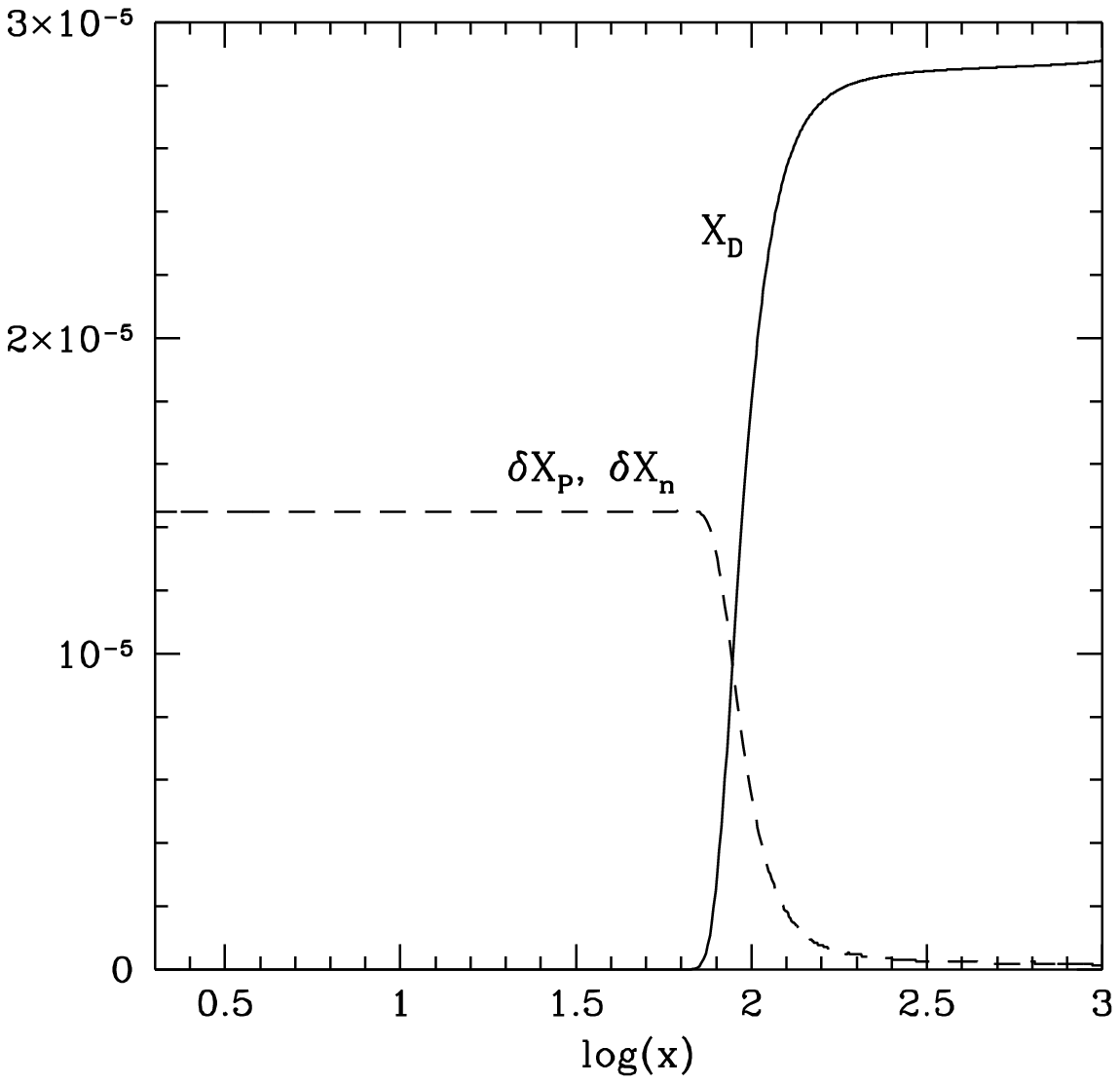,height=10truecm,width=10truecm}}}
\vspace{4.8cm}
\noindent {\small {\bf Fig. 4}: Destruction of  
deuterium ($X_D$), as matter is accreted. Protons 
and neutrons created because of this are also 
shown (dashed curve).}
\end{figure}

\end{document}